\documentclass[preprint,showpacs,aps]{revtex4} 
 
\usepackage{graphicx}
\begin{document}  

\title{Efficient Stochastic Simulations of Complex 
Reaction Networks on Surfaces} 
\author{Baruch Barzel and Ofer Biham}  
\affiliation{  
Racah Institute of Physics,   
The Hebrew University,   
Jerusalem 91904,   
Israel}  

 
\newcommand{\N}[1][]
{
\langle N^{#1} \rangle
}

\newcommand{\F}[1]
{
F_{\rm #1}
}

\newcommand{\W}[1]
{
W_{\rm #1}
}

\newcommand{\A}[1]
{
A_{\rm #1}
}

\newcommand{\f}[1]
{
f_{\rm #1}
}

\newcommand{\R}[1]
{
R_{\rm #1}
}

\newcommand{\Svis}
{
a_{\rm H}/W_{\rm H}
}

\newcommand{\Svac}
{
W_{\rm H}/f_{\rm H}
}

\newcommand{\Prob}[1]
{
P_{\rm #1}(N_{\rm #1})
}


\begin{abstract}  

Surfaces serve as highly efficient catalysts for a vast variety of
chemical reactions.
Typically, such surface reactions involve billions
of molecules which diffuse and react over macroscopic areas.
Therefore, stochastic fluctuations are negligible and
the reaction rates can be evaluated using rate equations,
which are based on the mean-field approximation.
However, in case that the surface is partitioned into a large number
of disconnected microscopic domains, the number of reactants in each
domain becomes small and it strongly fluctuates.
This is, in fact, the situation in the interstellar medium,
where some crucial reactions take place on the surfaces  
of microscopic dust grains.
In this case rate equations fail and
the simulation of surface reactions requires stochastic methods 
such as the master equation.
However, in the case of complex reaction networks, 
the master equation becomes infeasible 
because the number of equations proliferates exponentially.
To solve this problem, 
we introduce a stochastic method based on moment equations.
In this method
the number of equations is dramatically reduced to 
just one equation for each reactive species and one equation 
for each reaction.
Moreover, the equations can be easily constructed
using a diagrammatic approach.
We demonstrate the method for a set of astrophysically
relevant networks of increasing complexity.
It is expected to be applicable in many other contexts 
in which problems that exhibit analogous structure appear,
such as surface catalysis in nanoscale systems,
aerosol chemistry in stratospheric clouds 
and genetic networks in cells.

\end{abstract}  
  
\pacs{05.10.-a,82.65.+r,98.58.-w} 
 
\maketitle  

\section{Introduction}
\label{Introduction}

The catalysis of chemical reactions by surfaces plays a crucial 
role in a vast range of physical, chemical and biological systems.
In many cases, the surfaces are of macroscopic dimensions
and the reactants appear in large quantities.
Under these conditions, the law of large numbers applies
and fluctuations in the surface concentrations
of reactants and their reaction rates become negligible. 
As a result, these reactions can be analyzed using rate
equation models that incorporate the mean-field approximation
and ignore fluctuations.

Consider the case of a surface, which is partitioned 
into microscopic domains, that are disconnected, namely reactants
cannot diffuse between them.
The population of reactive atoms and molecules in
each domain becomes small and fluctuations become important.
As a result, rate equations fail and the simulation of these reactions
requires stochastic methods such as 
direct integration of the master equation 
\cite{vanKampen1981,Oppenheim1977,Karlin1998,Nasell2001}
or Monte Carlo (MC) simulations
\cite{Gillespie1977,Newman1999}.
While MC simulations require the accumulation
of statistical data over long times, the master equation
provides the probability distribution from which the reaction rates
can be obtained directly.
In certain cases, the master equation can be solved using a generating
function
\cite{McQuarrie1963,McQuarrie1964,Gardiner1985,Assaf2006,Assaf2007}.  
The set of coupled ordinary differential
equations is then transformed into a single partial differential equation
for the generating function. This equation can solved numerically
and in a few cases can also be solved analytically
\cite{Green2001,Biham2002}.
The master equation can also be approximated using the
Fokker-Planck equation
\cite{Montroll1946}.
This is a partial differential equation in which the
population sizes of the reactive species are represented
by continuous variables
\cite{Kolmogorov1931}.
The master equation and related 
methods described above are useful
for simple reaction networks, 
which involve few reactive species.
However, as the number of reactive species increases,
the number of equations 
in the master equation
quickly proliferates.
This makes
the master equation infeasible for complex 
networks including a large number of reactive species.

The recently introduced multiplane method 
provides a dramatic reduction in the number of equations
\cite{Lipshtat2004}.
In this method, the reaction networks are described by graphs, where
each node represents a species and each edge represents
a reaction. Typically, these networks are sparse, namely most 
species react only with few other species.
In the multiplane method one
breaks the network into a set of fully connected
subnetworks (cliques).
Lower-dimensional master equations are constructed for the
marginal probability distributions associated with the cliques,
with suitable couplings between them.
This enables the simulation of large networks much
beyond the feasibility limit of the master equation.
However, it turns out that the construction of the multiplane
equations for complex networks is difficult.

In this paper we present a method 
for stochastic simulations of surface reaction networks,
which is based on moment equations.
This method exhibits crucial advantages over the multiplane method
\cite{Barzel2006}.
The number of equations is further reduced to 
one equation for each reactive species
(node) 
and one equation for each reaction (edge). 
Thus, for typical sparse networks the complexity of the stochastic 
simulation becomes comparable to that of the rate 
equations.
Unlike the master equation (and the multiplane method) 
there is no need to adjust the cutoffs, namely
the same set of equations is used under all physical conditions.
Moreover, for any given network the moment equations
can be easily constructed using a diagrammatic approach.
This enables to automate the construction of the set of 
equations - a feature which is essential in the case of 
complex networks.
Furthermore, the moment equations are linear in terms of
the moments. Thus, the stability and convergence properties
can be easily controlled and the steady state solution can
be obtained by standard algebraic methods.

Below we consider some examples of surface reaction systems in
which fluctuations play an important role.
In surface catalysis systems, the surface is often partitioned 
into facets on nanometric dimensions, with little diffusion
between the facets.
Under these conditions the number of reactive atoms and molecules
residing on a facet is small and fluctuations are strong.
\cite{Suchorski1998,Suchorski1999,Suchorski2001,Johanek2004,Pineda2006,Liu2002}. 
Another important example of chemical reaction networks that
require stochastic analysis appears in the field of
interstellar chemistry.
Some chemical reactions in interstellar clouds take
place on the surfaces of dust grains
\cite{Spitzer1978,Hartquist1995,Hasegawa1992,Herbst1995}. 
These include molecular hydrogen formation 
\cite{Gould1963,Hollenbach1970,Hollenbach1971a,Hollenbach1971b}
as well as 
reaction networks that form ice mantles and certain 
organic molecules.
Unlike gas phase reactions in cold clouds that mainly produce 
unsaturated molecules
\cite{Herbst2005}, 
surface processes  
are dominated by hydrogen-addition reactions
that result in saturated, hydrogen-rich molecules, such as
water (H$_2$O),
ammonia (NH$_3$), methane (CH$_4$),
formaldehyde (H$_2$CO) and methanol (CH$_3$OH)
\cite{Hiraoka1998,Hidaka2004}. 
In particular, recent experiments have shown that methanol cannot
be efficiently produced by gas phase reactions
\cite{Geppert2006}.
On the other hand, there are indications that it can be 
efficiently produced on ice-coated grains
\cite{Watanabe2005}. 
Therefore, the ability  
to perform efficient simulations of chemical reactions on interstellar
grains is of great importance.

Due to the submicron size of the grains and the low
gas density, the populations of reactive species per
grain are small and strongly fluctuate.
Therefore, 
rate equations are not suitable
\cite{Tielens1982,Charnley1997,Caselli1998,Shalabiea1998}  
and stochastic methods such as
the direct integration of the master equation 
\cite{Biham2001,Green2001}
or MC simulations
\cite{Charnley2001}
are required.
As discussed above, 
these methods apply in the case of small networks
but become infeasible for large networks
\cite{Stantcheva2002,Stantcheva2003}.

Here we demonstrate the moment-equation method 
for a set of astrophysically relevant
networks of increasing complexity, culminating in the network that
describes the production of methanol. 
It is shown that in spite of the small number of equations and the
ease of their construction, the accuracy of the results is not
compromised. The results of the moment equations are in excellent
agreement with the master equation and coincide with the rate
equations in the limit of large grains.

The paper is organized as follows. 
In Sec. II we consider the production rate of molecular 
hydrogen using the rate equation, master equation and 
moment equation methods. 
In Sec. III we consider the simulation of complex
reaction networks using the moment equations. 
The construction of the set of moment equations using
diagrammatic methods is presented in Sec. IV, followed
by a summary in Sec. V.

\section{Molecular Hydrogen Formation}
\label{sec:H_production}

\subsection{Rate Equations}

Consider the reaction H$+$H$\longrightarrow$H$_2$
that takes place on the surfaces of interstellar dust
grains. Hydrogen atoms from the gas phase collide and
stick to the surface of a dust grain. They diffuse 
between adsorption sites on the surface and when two
of them encounter each other they recombine and form
an H$_2$ molecule, which desorbs into the gas phase.

For simplicity, we assume that the grains are spherical
and denote their diameter by $d$.  
The cross-section of a grain is 
$\sigma=\pi d^2/4$ 
and its surface area is $\pi d^2$. 
The density of adsorption sites on the surface of a grain 
is denoted by 
$s$ (sites cm$^{-2}$). 
Thus, the number of  
adsorption sites on the grain is 
$S=\pi d^2 s$. 
Each grain is exposed to a flux 
$F = n v \sigma$ (s$^{-1}$)
of H atoms,
where $n$ (cm$^{-3}$)
is the density of H atoms in the gas phase
and $v$ (cm s$^{-1}$)
is their average velocity, 
which is determined by the gas temperature.
It is also useful to define the flux 
in units of monolayers (ML) per second,
namely $f=F/S$ (ML s$^{-1}$).
Clearly, this flux is independent of the grain size.

The desorption rate of H atoms from the
grain is given by   
$W =  \nu \cdot \exp (- E_{1} / k_{B} T)$,    
where $\nu$ is the attempt rate   
(standardly taken to be $10^{12}$ s$^{-1}$),   
$E_{1}$   
is the activation energy barrier for desorption   
of an H atom,
and $T$ (K)
is the surface temperature.  
The hopping rate of adsorbed atoms between 
adjacent sites on the surface is 
$a =  \nu \cdot \exp (- E_{0} / k_{B} T)$,   
where
$E_{0}$ is the activation energy barrier for hopping 
of the atoms.  

Molecular hydrogen formation on astrophysically relevant
surfaces of silicates, carbon and ice 
has been studied by laboratory experiments
during the past decade
\cite{Pirronello1997a,Pirronello1997b,Roser2002,Hornekaer2003,Hornekaer2005,Vidali2005,Amiaud2006}. From the results of 
these experiments one can extract, for each surface, the 
energy barriers defined above and the density of adsorption sites.
In the simulations presented below, 
the density of adsorption sites was 
$s=5 \times 10^{13}$ (sites cm$^{-2}$), 
the activation energies were
$E_0 = 22$ meV for diffusion and
$E_1 = 32$ meV for desorption. 
These parameters are rounded values of the experimental results
for hydrogen recombination on silicates
\cite{Katz1999}.
The grain temperature was taken as $T=10$K.
Here we assume that diffusion occurs only by thermal hopping,  
in agreement with the experimental results  
\cite{Katz1999,Perets2005}.  
For small grains, it is convenient to replace the  
hopping rate $a$ (hops s$^{-1}$) by 
the sweeping rate
$A = a / S$, 
which is approximately the inverse of the time it takes  
for an H atom to visit nearly all  
the adsorption sites on the grain surface
(for a more precise evaluation of the sweeping rate see Refs. 
\cite{Krug2003,Lohmar2006}). 
The rate equation for this reaction takes the form

\begin{equation}
{d \langle N \rangle \over dt} = 
F - W \langle N \rangle 
- 2A\langle N \rangle ^2,
\label{eq:rate_1specie}
\end{equation}

\noindent
where $\langle N \rangle$ is the 
average population size of H atoms residing on the grain.
The first term on the right hand side of Eq. 
(\ref{eq:rate_1specie}) 
describes the incoming 
flux of H atoms.
The second term  
accounts for the desorption of H atoms from the
surface, which is proportional to the number of adsorbed atoms.
The third term  
accounts for the depletion 
in the population
of adsorbed H atoms
due to the recombination process. 

The production rate of H$_{2}$ molecules on a single grain,
is given by  

\begin{equation}
R =  A \langle N \rangle ^{2}. 
\label{eq:R_H2}
\end{equation}

\noindent
For simplicity we assume 
that these molecules desorb from the surface
upon formation. 

The rate equation 
(\ref{eq:rate_1specie})
can be solved either analytically 
or by numerical integration.
The steady state solution can be obtained by setting 
the left hand side of Eq. (\ref{eq:rate_1specie})
to zero. 
Under steady state conditions, the average population
size of H atoms on a grain is

\begin{equation}
\langle N \rangle = 
{1 \over 4} 
\left( 
W \over A 
\right)
\left[
-1 + \sqrt{1 + 8 \gamma}
\right]
\label{eq:rate_ss}
\end{equation}

\noindent
and the formation rate of molecules per grain is

\begin{equation} 
R =
{W^2 \over 16 A}
\left[
-1 + \sqrt{1 +  {8 \gamma}}
\right]^2, 
\label{eq:eta_ss}
\end{equation}

\noindent
where $\gamma = FA/W^2$.
Note that $\gamma$ is independent of the grain size.
Under these conditions, one can define
the recombination efficiency 

\begin{equation}
\eta = {R \over {F/2}}, 
\label{eq:eta}
\end{equation}

\noindent
which is the fraction of adsorbed atoms 
that come out in molecular form.
Note that 
$0 \le \eta \le 1$. 

One can identify two limits.
In the limit where
$\gamma \ll 1$
the desorption process is dominant 
and the recombination efficiency is low.
In this case,
the recombination rate per grain is
$R \simeq A F^2/W^2$
and the recombination efficiency is
$\eta \simeq 2 AF/W^2$.
Since $R$ depends quadratically on the flux,  
this regime is referred to as second order kinetics.
In the opposite limit,
where 
$\gamma \gg 1$,
the recombination process is dominant and the efficiency is 
$\eta \simeq 1$. 
This is the regime of
first order kinetics,
in which $R$ depends linearly on $F$
\cite{Biham1998,Biham2002}.
  
Under given flux and surface temperature, 
for grains that are large enough to hold many H atoms,
Eq. (\ref{eq:rate_1specie})
provides a good description of the recombination process.
However, in the limit of small grains and low flux,
$\langle N \rangle$ 
may be reduced to order unity or less.
Under these conditions,
Eq. (\ref{eq:rate_1specie})
becomes unsuitable, 
because it ignores the discrete nature of the population
of adsorbed atoms and its fluctuations.

\subsection{The Master Equation}

To account correctly for the recombination rate on small grains,
simulations using the master equation are required.
The dynamical variables
of the master equation are the probabilities
$P(N)$
of having a population of $N$ hydrogen atoms 
on the grain.
In the case of hydrogen recombination, 
the master equation takes the form

\begin{eqnarray}
\dot {P} (N ) &=& F  
\left[ P (N -1) - P (N ) \right]
+ W  \left[ (N +1) P 
(N +1) - N  P (N ) \right] 
\nonumber \\
&+& A  [ (N +2) (N +1) P (N +2) 
- N (N -1) P (N ) ],
\label{eq:master_1specie}
\end{eqnarray}

\noindent
where $N=0,1,2,\dots$.
The first term on the right hand side of 
Eq. (\ref{eq:master_1specie})
describes the effect of the incoming flux. 
The probability $P(N )$ increases when an H atom is adsorbed by
a grain that already has $N -1$ adsorbed H atoms, and decreases when 
it is adsorbed on a grain with $N$ atoms.
The second term accounts for the desorption process. 
The third term describes the recombination process. 
The recombination rate is proportional to the number of pairs of
H atoms on the grain, namely $N (N -1) / 2$.
Therefore, the H$_2$ production rate can be expressed in terms
of the moments of $P(N)$ as 

\begin{equation}
R = A ( \langle N^2 \rangle -  \langle N \rangle ).
\label{eq:master_production}
\end{equation}

\noindent
In numerical simulations the master equation must be truncated in order 
to keep the number of equations finite.
A convenient way to achieve this is to assign an upper cutoff 
$N^{\rm max}$ on the population size.
The number of equations is thus $N^{\rm max} + 1$.
The truncated master equation is valid if the probability 
to have a larger population than the assigned cutoff is 
vanishingly small. 
Therefore, the upper cutoff should be chosen according to
the parameters of the simulation.

\subsection{The Moment Equations}
\label{sec:moment}

As noted above, 
the population size of the adsorbed H atoms
is given by the first moment
of $P(N)$, 
while
the reaction rate can be expressed in terms of the 
difference between the second moment and the first moment.
Therefore, a closed set of equations for the time derivatives of
these first and second moments could directly provide all the information
that we need about the population size and reaction rates
\cite{Lipshtat2003}.
Such equations can be obtained from the master equation using
the identity

\begin{equation}
{d {\langle N^k \rangle} \over dt} = 
\sum_{N=0}^{N^{\rm max}} 
N^k \dot P(N),
\end{equation}
 
\noindent
where $k$ is an integer.
Inserting $\dot P(N)$ according to 
Eq. (\ref{eq:master_1specie}),
one obtains the
moment equations 

\begin{eqnarray}
{d \langle N \rangle \over {dt} } &=& 
F - W \langle N \rangle 
- 2 A ( \langle N^2 \rangle  - \langle N \rangle)  
\nonumber \\
{d \langle N^2 \rangle \over {dt} } &=& 
F+( 2 F+ W-4A ) \langle N \rangle
+(8 A - 2 W) \langle N^2 \rangle
- 4 A \langle N^3 \rangle. 
\label{eq:momentsH}
\end{eqnarray}

\noindent
This is a set of coupled differential equations,
which are linear in the moments
$\langle N^k \rangle$.
Although we have written the equations only for the 
first two moments, 
the right hand side includes the third moment
for which we have no equation.
In order to close the set of equations one must express the
third moment in terms of the first two moments.
Different such expressions have been proposed.
For example, in the context of birth-death processes
the relation 
$\langle N^3 \rangle = 
\langle N^2 \rangle \langle N \rangle$
was used 
\cite{McQuarrie1967}.
This choice makes the moment equations 
nonlinear, with possible effects on their 
stability.
Another common choice is to
assume that the third central moment is 
zero (which is exact for symmetric distributions)
and use this relation in order to express
the third moment in terms of the first and second moments
\cite{Gomez-Uribe2007}.

Here we set up the closure condition
by imposing a highly restrictive cutoff 
on the master equation.
The cutoff is set at $N^{\rm max} = 2$.
This is the minimal cutoff 
that still enables the recombination process to take place.
Under this cutoff, one obtains the following relation
between the first three moments
\cite{Lipshtat2003} 

\begin{eqnarray}
\langle N^3 \rangle = 3\langle N^2 \rangle 
-2 \langle N \rangle.
\label{eq:third_moment}
\end{eqnarray}

\noindent
Using this result, one can bring the moment equations
[Eq. (\ref{eq:momentsH})]
into a closed form:

\begin{eqnarray}
{{d \langle N \rangle} \over dt} &=& 
F - W \langle N \rangle -
2A (\langle N^2 \rangle - \langle N \rangle)
\nonumber \\
{{d \langle N^2 \rangle} \over dt} &=& 
F +(2F + W + 4A ) \langle N \rangle -
(2W + 4A) \langle N^2 \rangle.
\label{eq:moment_2}
\end{eqnarray}

\noindent
Numerical integration of these equations provides the first two moments,
from which the population size and reaction rate are obtained.
The steady state solution of the moment 
equations can be obtained by setting the derivatives
on the left hand side of 
Eq. (\ref{eq:moment_2}) 
to zero. 
One obtains

\begin{eqnarray}
\N &=& {{F(A + W)} \over {2AF + WA + W^2}} \nonumber \\
\N[2] &=& {{F(F + A + W)} \over {2AF + WA + W^2}}. 
\label{eq:moment_ss}
\end{eqnarray}  

\noindent
Using 
Eq. (\ref{eq:master_production})
we find that the 
recombination rate 
is

\begin{equation}
R = {{f^2} \over {W}} 
\left[ 
{{S^2} \over {1 + {S \over {a/W}} + {S \over {W/f}}}}
\right].
\label{eq:R_moment_ss}
\end{equation} 

\noindent
In this system one can identify two
characteristic spatial scales,
namely 
$S_1 = a/W$ 
and 
$S_2 = W/f$
\cite{Lipshtat2003}.
The limit of small grains is obtained when
$S$ is smaller than both $S_1$ and $S_2$.
In this limit the populations size of atoms
on the grain is small, the rate equation
fails while the moment equations are accurate.
The limit of large grains is obtained when
$S$ is larger than both 
$S_1$ and $S_2$.
In this limit the population size of atoms
on the grain is large,
the rate equation applies
and the reaction rate
satisfies
$R \propto S$.
Since $F = fS$ and $A = a/S$,
the limit of large grains can be
expressed by

\begin{equation}
F \gg W \gg A.
\label{eq:large_grain}
\end{equation}

We recall that 
the moment equations were derived
by imposing a strict cutoff on the master equation.
One may thus expect that the equations 
will apply only in the case of small grains, 
where this cutoff is suitable.
However, it turns out that the moment equations 
are valid much beyond this limit.
Below we show that the reaction rate obtained from
the moment equations is accurate for both small
and large grains.

\begin{figure}
\includegraphics[width=5.0in]{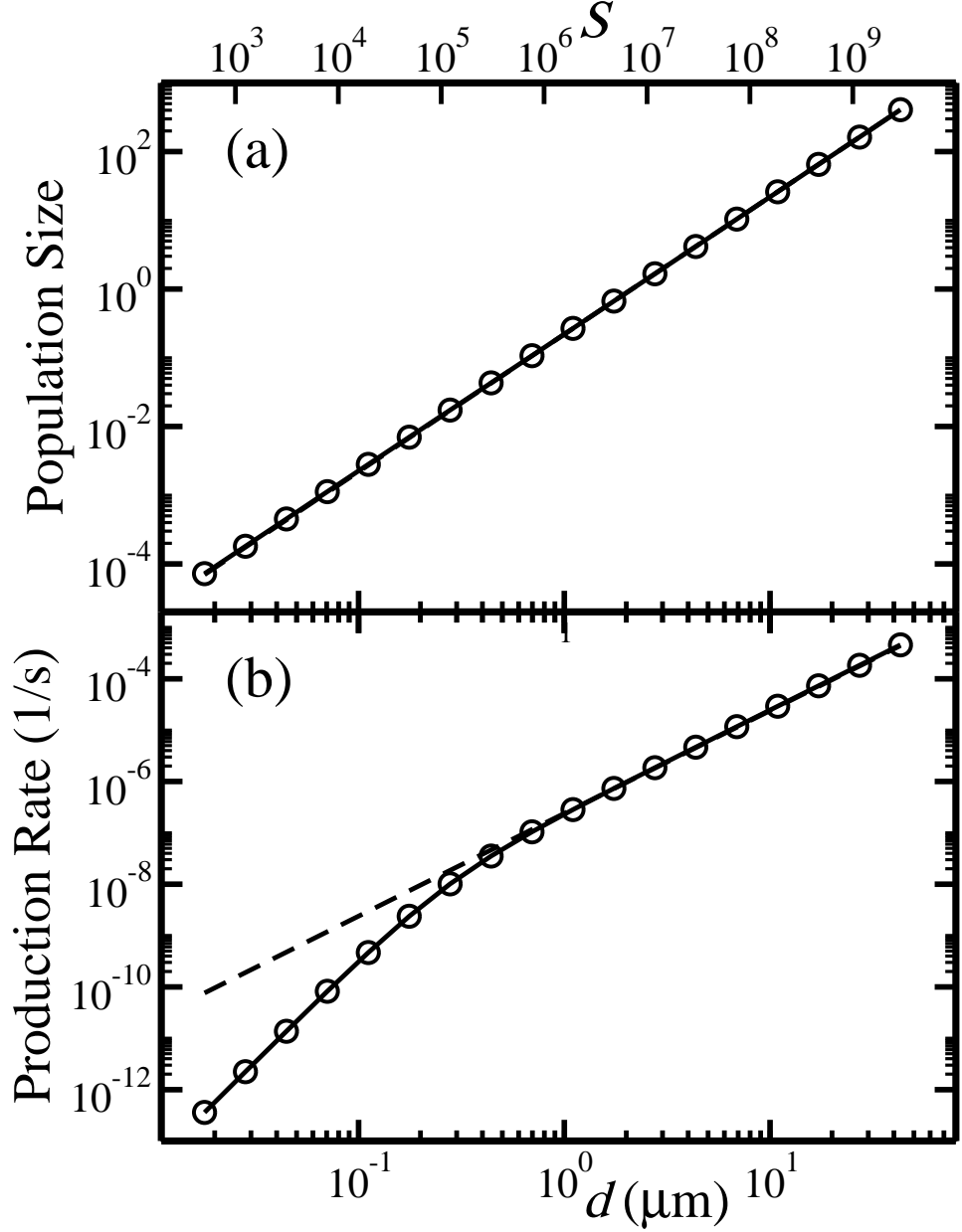}
\caption{
The population size of H atoms (a) and the
production rate of H$_2$ molecules (b) on the surface of a grain
vs. grain diameter, $d$,
under conditions of second-order kinetics.
The number of adsorption sites, $S$, is also shown.
The results of the moment equations (circles)
are in perfect agreement with the master equation 
(solid line) for all grain sizes.
In the limit of large grains they also coincide with  
the rate equation (dashed line).
}
\label{fig1}
\end{figure}

\begin{figure}
\includegraphics[width=5.0in]{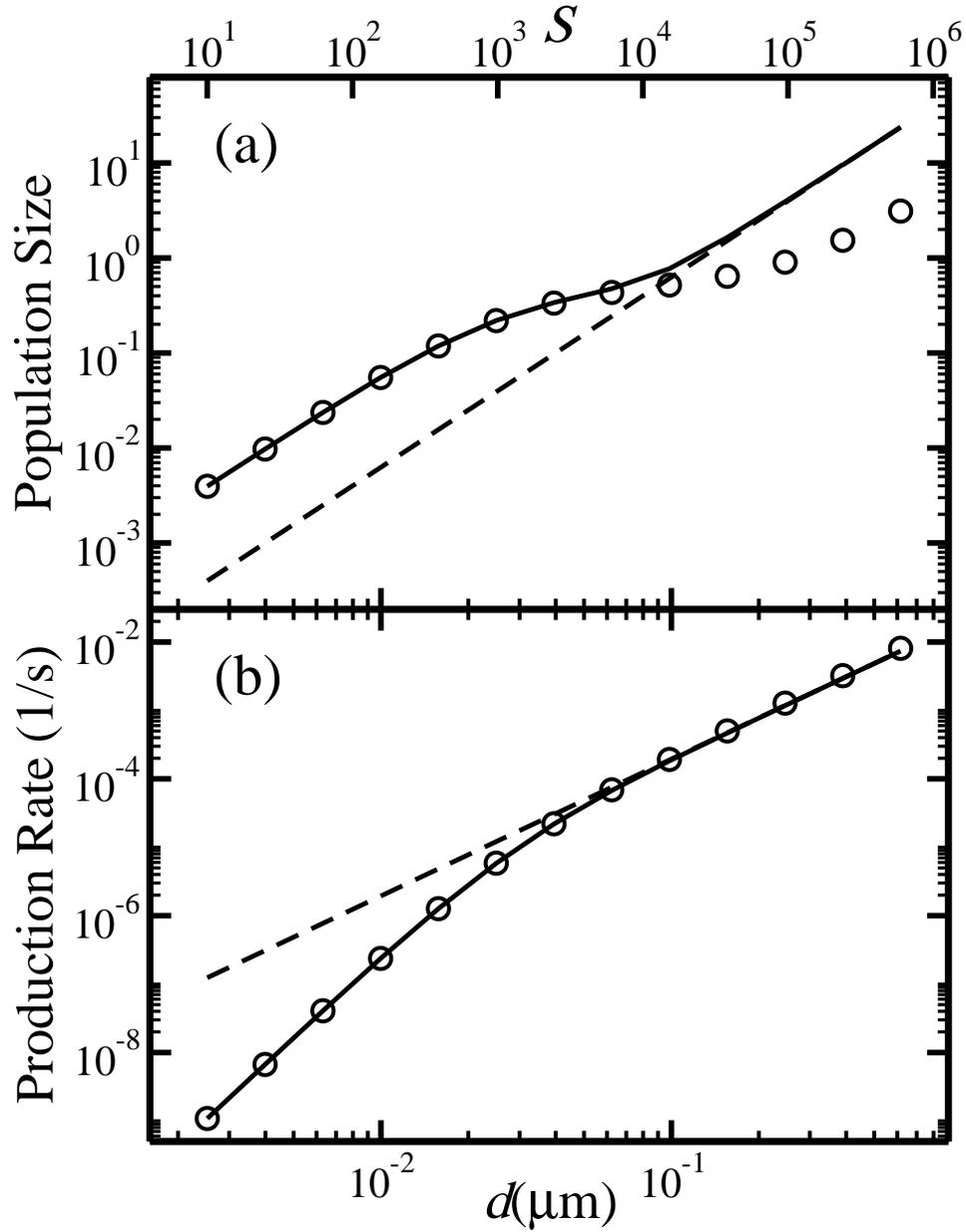}
\caption{
The population size of H atoms (a) and the
production rate of H$_2$ molecules (b) on the surface of a grain
vs. grain diameter, $d$,
under conditions of first-order kinetics.
The production rate obtained by
the moment equations (circles)
is in perfect agreement with the master equation 
(solid line) for all grain sizes.
However, the population size deviates for large grains.
The rate equation results are also shown (dashed lines).
}
\label{fig2}
\end{figure}

In Fig. \ref{fig1} we present the 
population size of H atoms and the
reaction rate, vs. grain diameter, 
obtained by the moment equations (circles).
The flux of H atoms was 
$F = 1 \times 10^{-11}S$ (atoms s$^{-1}$). 
The parameters used in 
Fig. \ref{fig1} 
satisfy the conditions for second-order kinetics, 
namely
$\gamma \ll 1$.
The results are in perfect agreement with the master equation (solid lines)
for the entire range of grain sizes.
For small grains, the rate equation (dashed lines)
over-estimates the recombination rate, but coincides
with the master equation and the moment equations for large grains. 
In Fig. \ref{fig2} 
we present the population size of H atoms and the 
reaction rate, vs. grain diameter,
under the conditions of first-order kinetics, where
$\gamma \gg 1$.
These conditions were obtained by increasing the flux to
$F = 2.75 \times 10^{-8}S$ (atoms s$^{-1}$).
The moment equations still
provide accurate results for the reaction rate 
on grains of all sizes.
However, for large grains the population size 
of adsorbed H atoms, obtained from the moment equations,
deviates from the master equation results.

In conclusion, one can identify 
two kinetic regimes (first and second order)
and two limits (small and large grains).
In the limit of small grains, the moment equations
are valid, as expected, for both kinetic regimes
(Table I).
In the case of second order kinetics, the moment
equations are valid both for small grains and
for large grains.
They provide accurate results both for the 
average number of atoms on a grain and for for
the reaction rate.
In the case of large grains under conditions of
first order kinetics the situation is different.
The moment equations still provide accurate results
for the reaction rate but are not suitable for the
evaluation of the population size 
on a grain.
In the next section we present a more complete
analysis of the moment equations and their validity.

\subsection{The Validity of the Moment Equations}
\label{chapter2a}

The moment equations were derived on the basis of a 
very strict cutoff, allowing at most two atoms
to simultaneously reside on the surface of the grain.
Nevertheless, it
turns out that the equations apply even under conditions
in which the population size of adsorbed atoms is large.
Here we examine the validity of the 
moment equations in the regimes of first and second order 
kinetics and in the limits of small and large grains.
For sufficiently small grains the population 
of adsorbed atoms on a grain is small and is consistent with
the strict cutoff imposed in the derivation of the moment equations.
Therefore in the limit of small grains the approximation of 
Eq. (\ref{eq:third_moment}) 
is justified, and the moment 
equations are valid.

In the limit of large grains
the population of adsorbed atoms on a grain is large and
the rate equation becomes accurate.
Therefore, we will test the validity of the
moment equations in this limit by comparison with
the rate equations.
In the limit of large grains 
the relations  
$F \gg W$ and $W \gg A$
are satisfied.
We will first consider the case of second order kinetics,
where $FA \ll W^2$.
Using the relations above, we find that
Eqs. 
(\ref{eq:moment_ss})
are reduced to
$\langle N \rangle \simeq F/W$
and 
$\langle N^2 \rangle - \langle N \rangle \simeq (F/W)^2$.
Thus the results of the moment equations for the
population size and for the reaction rate coincide
with the rate equations.

In the case of first order kinetics
$FA \gg W^2$.
In the limit of large grains, we obtain from
Eq. (\ref{eq:moment_ss})
that
$\langle N^2 \rangle - \langle N \rangle \simeq F/(2A)$,
which is consistent with the H$_2$ production rate obtained
from rate equations.
On the other hand, the expression for
the first moment is reduced to
$\langle N \rangle \simeq W/(2A)$,
which is inconsistent with the rate equations,
in which 
$\langle N \rangle \simeq \sqrt{F/(2A)}$.
These results are in line with the numerical results
shown above, indicating that the moment equations
are suitable for the evaluation of reaction rates
on large grains both in first and second order kinetics.
However, they provide the correct 
population size of atoms on large grains only in the
case of second order kinetics and not in first order
kinetics.

\begin{figure}
\includegraphics[width=4.5in]{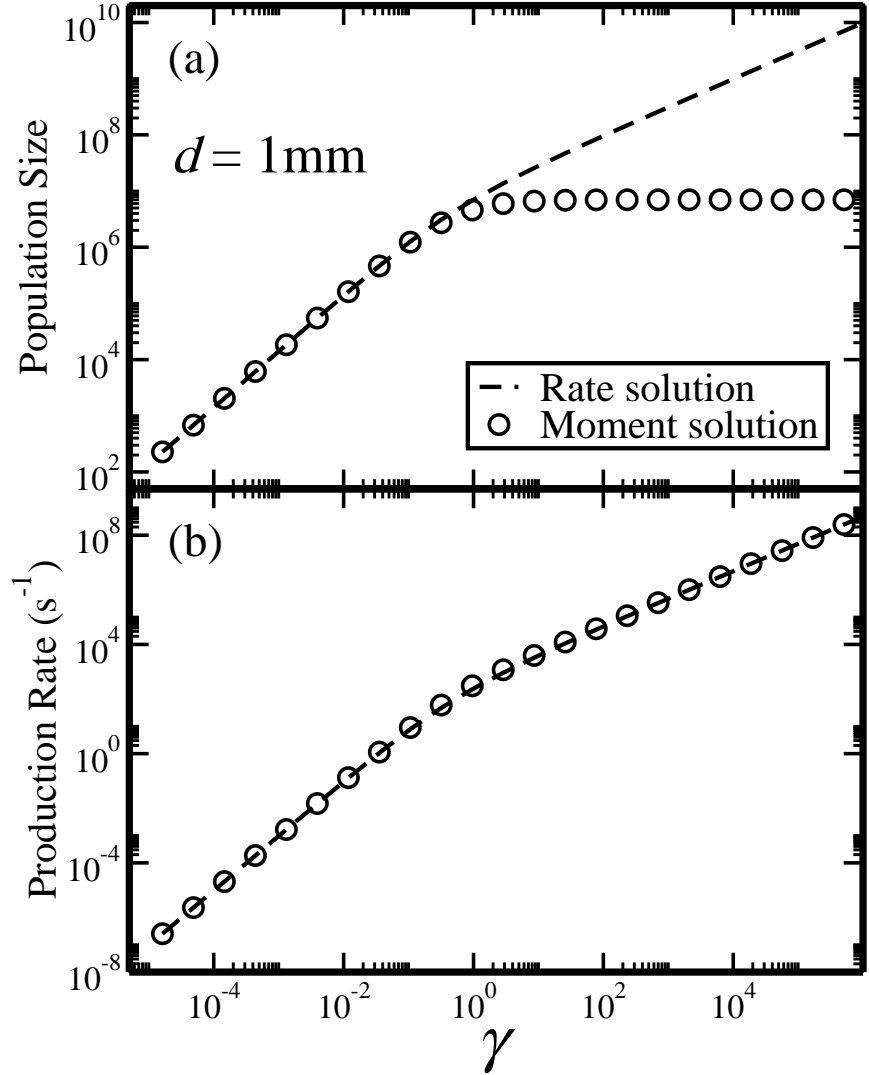}
\caption{
(a)
The population size of H atoms per grain vs.
the parameter
$\gamma$,
as obtained from the 
moment equations for a very large grain of diameter
$d = 1$mm (circles).
Under these conditions
the rate equation is accurate.
For $\gamma < 1$
(second order kinetics), the moment equations are 
in perfect agreement with the rate equation (dashed line).
This is in spite of the large population size. 
For $\gamma > 1$
(first order kinetics), 
the moment 
equations do not provide accurate results for the 
population size; 
(b)
The production rate of H$_2$ molecules per grain vs.
$\gamma$,
as obtained from the moment equations 
(circles).
The results are in perfect agreement with those of 
the rate equation (dashed line) both for second order 
and for first order kinetics (even when there is a
deviation in the first moment).
}
\label{fig3}
\end{figure}

In Fig. \ref{fig3}(a) we show the first moment,
$\N$,
as obtained from the moment equations (circles) vs. 
$\gamma$,
for a very large grain of diameter 
$d = 1$mm, where
$S \approx 1.6 \times 10^{12}$. 
The results are compared with those of the rate equation (dashed line),
which under these conditions is accurate. 
For
$\gamma \ll 1$
(second order kinetics)
the rate equation and the moment equations
coincide.
For
$\gamma \gg 1$ 
(first order kinetics) 
the moment-equation results 
for the average population size
deviate from those of 
the rate equation.
Note that even when
$\gamma \ll 1$,
where the moment equations apply, 
the population size of atoms per grain 
is very large, much beyond the cutoffs used
in the construction of the equations.
In 
Fig. \ref{fig3}(b) 
we present the production rate of 
H$_2$ molecules per grain vs.
$\gamma$,
as obtained from the moment equations (circles).
The results are in perfect agreement with those 
of the rate equations (dashed line)
in the regime of first order kinetics
as well as in the regime of
second order kinetics.
These results can be generalized to more complex 
networks. The analysis becomes more tedious, but
the conclusions about the domain of validity of
the moment equations remain the same. 
From a physical perspective, 
in first order kinetics the reaction is dominant
and desorption is suppressed, while in second order 
kinetics desorption is dominant and the reaction
rate is low and diffusion limited.

\section{Complex Reaction Networks}
\label{Complex}

\subsection{The Water-Producing Network}

To generalize the analysis beyond the case of a single
species, consider a simple chemical network that
involves three reactive species: H and O atoms and OH molecules
\cite{Caselli1998,Shalabiea1998,Stantcheva2001}.
For simplicity we denote the reactive species by
$X_1=$ H,
$X_2=$ O,
$X_3=$ OH,
and the resulting non-reactive species by
$X_4=$ H$_2$,
$X_5=$ O$_2$,
$X_6=$ H$_2$O.
The reactions that take place in this network include
H + O 
$\rightarrow$ 
OH
($X_1 + X_2 \rightarrow X_3$),
H + H 
$\rightarrow$ 
H$_2$
($X_1 + X_1 \rightarrow X_4$),
O + O 
$\rightarrow$ 
O$_2$
($X_2 + X_2 \rightarrow X_5$),
and
H + OH 
$\rightarrow$ 
H$_2$O
($X_1 + X_3 \rightarrow X_6$).
A graph of that network is displayed in Fig.
\ref{fig4}(a).
  
Now, the desorption rates of atomic and molecular species on the
grain are given by   
$W_i =  \nu \cdot \exp [- E_{1}(i) / k_{B} T]$,    
where $E_{1}(i)$   
is the activation energy barrier for desorption   
of species
$X_i$
and $T$ (K)
is the surface temperature.  
The hopping rate of adsorbed atoms between 
adjacent sites on the surface is 
$a_i =  \nu \cdot \exp [- E_{0}(i) / k_{B} T]$,   
where
$E_{0}(i)$ is the activation energy barrier for hopping 
of $X_i$ atoms (or molecules).   
The sweeping rate will be 
$A_i = a_i / S$. 

The master equation provides the time 
derivatives of the probabilities
$P(N_1,N_2,N_3)$
that a randomly chosen grain will have $N_i$ atoms or molecules 
of the reactive species $X_i$.
It takes the form

\begin{eqnarray}  
&&\dot P(N_1,N_2,N_3) = 
\sum_{i=1}^3  
F_i \left[ P(..,N_i-1,..) - P(N_1,N_2,N_3) \right]  
\nonumber \\ 
&&+\sum_{i=1}^3  
W_i \left[ (N_i+1) P(..,N_i+1,..) - N_i P(N_1,N_2,N_3) \right]  
\nonumber \\ 
&&+ \sum_{i=1}^2  
A_i [ (N_i+2)(N_i+1) P(..,N_i+2,..) 
- N_i(N_i-1) P(N_1,N_2,N_3) ]  
\nonumber\\ 
&&+ (A_1+A_2) [ (N_1+1)(N_2+1) P(N_1+1,N_2+1,N_3-1) 
- N_1 N_2 P(N_1,N_2,N_3) ] 
\nonumber \\ 
&&+ (A_1+A_3) [ (N_1+1)(N_3+1) P(N_1+1,N_2,N_3+1) 
- N_1 N_3 P(N_1,N_2,N_3) ]. 
\label{eq:Master}  
\end{eqnarray}  
 
\noindent  
The terms in the first sum describe the incoming flux,  
where $F_i$ (atoms s$^{-1}$)   
is the flux {\em per grain}  
of the species $X_i$.  
The second sum describes the effect of desorption.   
The third sum describes the effect of diffusion-mediated reactions 
between two identical atoms 
and the last two terms account for reactions between different species.
The rate of each reaction is proportional to the number of pairs of atoms
of the two species involved, and to the sum of their sweeping rates.
The average population size    
of the $X_i$ species on the grain is  
$\langle N_{i}\rangle = \sum_{N_1,N_2,N_3} N_i \ P(N_1,N_2,N_3)$,  
where 
$N_i=0,1,2 \dots$,
and
$i=1$, 2 or 3. 
The production rate 
per grain 
$R(X_k)$ 
(molecules s$^{-1}$)
of $X_k$ molecules
produced by the reaction
$X_i + X_j \rightarrow X_k$  
is given by 
$R(X_k) = (A_i+A_j) \langle N_i \  N_j \rangle$,
or by
$R(X_k)=A_i \langle N_i(N_i-1) \rangle$
in case that $i=j$.

For complex networks with more than one species 
the truncation of the master equation demands
setting upper cutoffs for all the reactive species,
$N_i^{\max}$, $i=1,\dots,J$, 
where $J$ is the number of reactive species. 
The number of coupled equations is then
$N_E = \prod_{i=1}^J (N_i^{\max}+1)$,
which grows exponentially 
with the number of reactive species.
This severely limits the applicability of the master equation
to interstellar chemistry
\cite{Stantcheva2002,Stantcheva2003}.
To reduce the number of equations one tries to use the lowest
possible cutoffs under the given conditions. 
In any case, to enable all reaction processes to take place,
the cutoffs must satisfy
$N_i^{\rm max} \ge 2$ 
for 
species that form homonuclear diatomic molecules
(H$_2$, O$_2$, etc.)
and
$N_i^{\rm max} \ge 1$
for other species.

As in the case of hydrogen recombination, 
the population sizes and reaction rates are completely 
determined by all the first moments and selected second
moments of the distribution
$P(N_1,N_2,N_3)$.
Therefore, a closed set of equations for the time derivatives of
these first and second moments could provide the 
complete information
on the population sizes and reaction rates.
For the simple network considered here one needs equations
for the time derivatives of the first moments
$\langle N_i \rangle$, $i=1,\dots,3$
and of the second moments
$\langle N_1^2 \rangle$,
$\langle N_2^2 \rangle$,
$\langle N_1 N_2 \rangle$
and
$\langle N_1 N_3 \rangle$.
We obtain these equations from the master equation using
the identity

\begin{equation}
{d {\langle N_1^aN_2^bN_3^c \rangle} \over dt} = 
\sum_{N_1,N_2,N_3=0}^{N_i^{\rm max}} 
N_1^aN_2^bN_3^c \dot P(N_1,N_2,N_3),
\label{eq:moment_abc}
\end{equation}

\noindent
where $a,b,c$ are integers.
Here we show three of the resulting moment equations:

\begin{eqnarray}
\frac{d \langle N_{1} \rangle }{dt} &=&
F_{1} - W_{1} \langle N_{1} \rangle 
-2A_{1} (\langle N_{1}^2 \rangle - \langle N_{1} \rangle)  
-(A_{1}+A_{2}) \langle N_{1}N_{2} \rangle 
-(A_{1}+A_{3}) \langle N_{1}N_{3} \rangle 
\label{eq:momentH}
\nonumber
\\
\frac{d \langle N_{1}^2 \rangle }{dt} &=&
F_{1}+(2F_{1}+W_{1}-4A_{1}) \langle N_{1} \rangle 
-(2W_{1} - 8A_{1}) \langle N_{1}^2 \rangle   
-4A_{1} \langle N_{1}^3 \rangle 
\nonumber\\
&+&(A_{1}+A_{2}) (\langle N_{1}N_{2} \rangle 
-2 \langle N_{1}^2N_{2} \rangle) 
+(A_{1}+A_{3}) (\langle N_{1}N_{3} \rangle 
-2 \langle N_{1}^2N_{3} \rangle) 
\label{eq:momentH2}
\nonumber
\\
\frac{d \langle N_{1}N_{3} \rangle }{dt} &=&
F_{1} \langle N_{3} \rangle +F_{3} \langle N_{1} \rangle 
-(W_{1}+W_{3}-3A_{1}-A_{3}) \langle N_{1}N_{3} \rangle 
-(3A_{1}+A_{3}) \langle N_{1}^2N_{3} \rangle 
\nonumber\\
&-&(A_{1}+A_{3}) \langle N_{1}N_{3}^2 \rangle 
-(A_{1}+A_{2}) (\langle N_{1}N_{2} \rangle - \langle N_{1}^2N_{2} \rangle 
+ \langle N_{1}N_{2}N_{3} \rangle). 
\label{eq:momentH_H2O}
\end{eqnarray}

\noindent
As expected, the right hand sides of these equations include
third order moments for which we have no equations.
If we add equations for their time derivatives,
they will include fourth order moments, and again will
not enable us to close the set of equations.
In order to close the set of moment equations
we must express the third order moments in terms of 
the first and second order moments.
We recall that in the case 
of a single reactive specie
this was achieved 
by imposing an extremely restrictive cutoff
on the master equation.
Generalizing it to our present system we
choose the minimal cutoffs that do not terminate any
of the reactions taking part in the network.
These cutoffs allow at most two atoms or molecules to be
adsorbed on the surface at any given time. 
Furthermore, any two atoms or molecules 
that reside simultaneously on the surface
must belong to species that react
with each other.
In our network
these cutoffs allow only eight non-vanishing probabilities,
namely,
$P(0,0,0)$,
$P(0,0,1)$,
$P(0,1,0)$,
$P(1,0,0)$,
$P(2,0,0)$,
$P(0,2,0)$,
$P(1,1,0)$
and
$P(1,0,1)$.
Expressing the third moments, with these cutoffs,
leads to the following results:

\begin{eqnarray}
\langle N_i^3 \rangle &=& 3 \langle N_i^2 \rangle 
- 2 \langle N_i \rangle
\nonumber \\
\langle N_i^2N_j \rangle &=& 
\langle N_iN_j \rangle
\nonumber \\
\langle N_iN_jN_k \rangle &=& 0.
\label{eq:barzel_rules}
\end{eqnarray}

\noindent
Using these rules to modify Eqs.
(\ref{eq:momentH_H2O})
one obtains a closed set of the form

\begin{eqnarray}
\frac{d \langle N_{1} \rangle }{dt} &=&
F_{1}-W_{1} \langle N_{1} \rangle 
-2A_{1} (\langle N_{1}^2 \rangle - \langle N_{1} \rangle )  
-(A_{1}+A_{2}) \langle N_{1}N_{2} \rangle 
-(A_{1}+A_{3}) \langle N_{1}N_{3} \rangle 
\label{eq:momentH_s}
\nonumber
\\
\frac{d \langle N_{2} \rangle }{dt} &=&
F_{2} - W_{2} \langle N_{2} \rangle 
-2A_{2} (\langle N_{2}^2 \rangle - \langle N_{2} \rangle)    
-(A_{1}+A_{2}) \langle N_{1}N_{2} \rangle  
\label{eq:momentO_s}
\nonumber
\\
\frac{d \langle N_{3} \rangle }{dt} &=&
F_{3} - W_{3} \langle N_{3} \rangle 
-(A_{1}+A_{3}) \langle N_{1}N_{3} \rangle  
+(A_{1}+A_{2}) \langle N_{1}N_{2} \rangle  
\label{eq:momentOH_s}
\nonumber
\\
\frac{d \langle N_{1}^2 \rangle }{dt} &=&
F_{1}+(2F_{1}+W_{1}+4A_{1}) \langle N_{1} \rangle 
-(2W_{1} +4A_{1}) \langle N_{1}^2 \rangle   
\nonumber\\
&-&
(A_{1}+A_{2}) \langle N_{1}N_{2} \rangle 
-(A_{1}+A_{3}) \langle N_{1}N_{3} \rangle 
\label{eq:momentH2_s}
\nonumber
\\
\frac{d \langle N_{2}^2 \rangle }{dt} &=&
F_{2}+(2F_{2}+W_{2}+4A_{2}) \langle N_{2} \rangle 
-(2W_{2}+4A_{2}) \langle N_{2}^2 \rangle  
-(A_{1}+A_{2}) \langle N_{1}N_{2} \rangle 
\label{eq:momentO2_s}
\nonumber
\\
\frac{d \langle N_{1}N_{2} \rangle }{dt} &=&
F_{1} \langle N_{2} \rangle +F_{2} \langle N_{1} \rangle 
-(W_{1}+W_{2}+A_{1}+A_{2}) \langle N_{1}N_{2} \rangle 
\label{eq:momentO_H_s}
\nonumber
\\
\frac{d \langle N_{1}N_{3} \rangle }{dt} &=&
F_{1} \langle N_{3} \rangle +F_{3} \langle N_{1} \rangle 
-(W_{1}+W_{3}+A_{1}+A_{3}) \langle N_{1}N_{3} \rangle. 
\label{eq:momentH_H2O_s}
\end{eqnarray}

\noindent
This set of equations is the minimal set required for
the system. It includes one equation that accounts for
the population size of each reactive specie
and one equation that accounts for the reaction rate
of each reaction.
As in the hydrogen system,
the equations were derived using extremely strict
cutoffs, that are expected to apply only in the limit of
small grains and low flux. 
Nevertheless, 
the moment equations are valid well beyond
the restrictive cutoffs under which they were derived.
In Fig. 
\ref{fig5}(a)
we present the population sizes of H ($\times$) and O ($+$) atoms
on a grain vs. grain diameter,
obtained from the moment equations.
In Fig. 
\ref{fig5}(b)
we show the production rates of
H$_2$ (circles),
O$_2$ (triangles)
and
H$_2$O (squares),
obtained from the moment equations.
The results are in excellent agreement with the master
equation and in the limit of large grains they also
coincide with the rate equations.
The parameters used in the simulations are
$F_1 = 5.0 \times 10^{-10}S$ (atoms s$^{-1}$),
$F_2 = 0.1F_1$ 
and
$F_3=0$.
The activation energies for diffusion and desorption were taken as
$E_0(1) = 44$, 
$E_1(1)=52$, 
$E_0(2) = 47$,
$E_1(2)=54$, 
$E_0(3)=47$ 
and 
$E_1(3)=54$ meV 
for H, O and OH respectively.
The surface temperature of the grain was $T=15$K.
For the species other than H there 
are no concrete experimental results,
and the values reflect the tendency 
of heavier species to bind
more strongly.

The generalization to more complex networks is straightforward.
As explained above, the cutoffs imposed on the master equation
give rise to a minimal set of non-vanishing probabilities,
that allow all the reactions to take place.
These non-vanishing probabilities are 
$P(0,\dots,0,N_i=1,0,\dots,0)$ 
for the each node, $i$,
and 
$P(0,\dots,0,N_i=1,0,\dots,0,N_j=1,0,\dots,0)$ 
for each edge connecting nodes $i$ and $j$.
A loop connected to the node $i$ corresponds to the
probability $P(0,\dots,0,N_i=2,0,\dots,0)$. 
All other probabilities vanish.
In the realm of two-body reactions additional probabilities
such as
$P(3,0,\dots,0)$ or $P(2,1,0,\dots,0)$ 
are never required.
Under these restrictions, one can easily show
that all the third moments that appear in the 
moment equations can be expressed in terms of  
second moments according to 
Eq. (\ref{eq:barzel_rules}).

\subsection{The Methanol Network}

\begin{figure}
\includegraphics[width=5.5in]{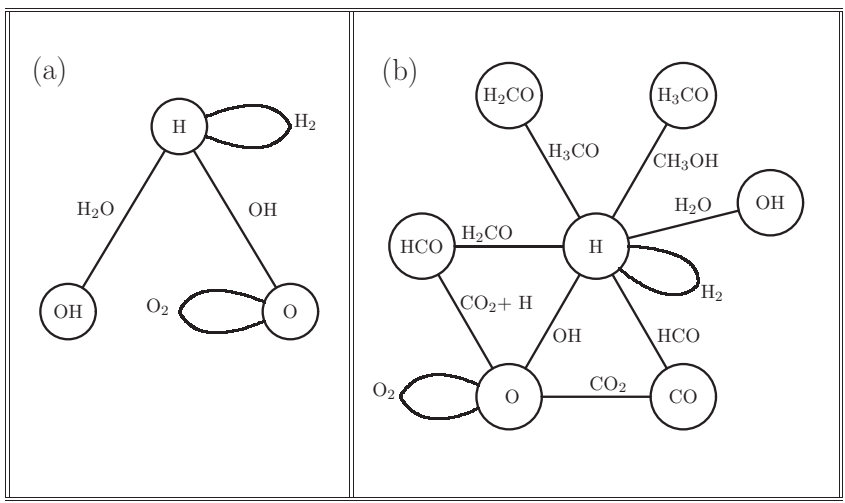}
\caption{
(a) The reaction network on a grain exposed to fluxes 
of H and O atoms and OH molecules.
Each node represents a reactive species and each edge represents
a reaction. The reaction products are specified near the edges;
(b) The network obtained by adding CO molecules, which gives rise
to the production of methanol.
}
\label{fig4}
\end{figure}

\begin{figure}
\includegraphics[width=5.0in]{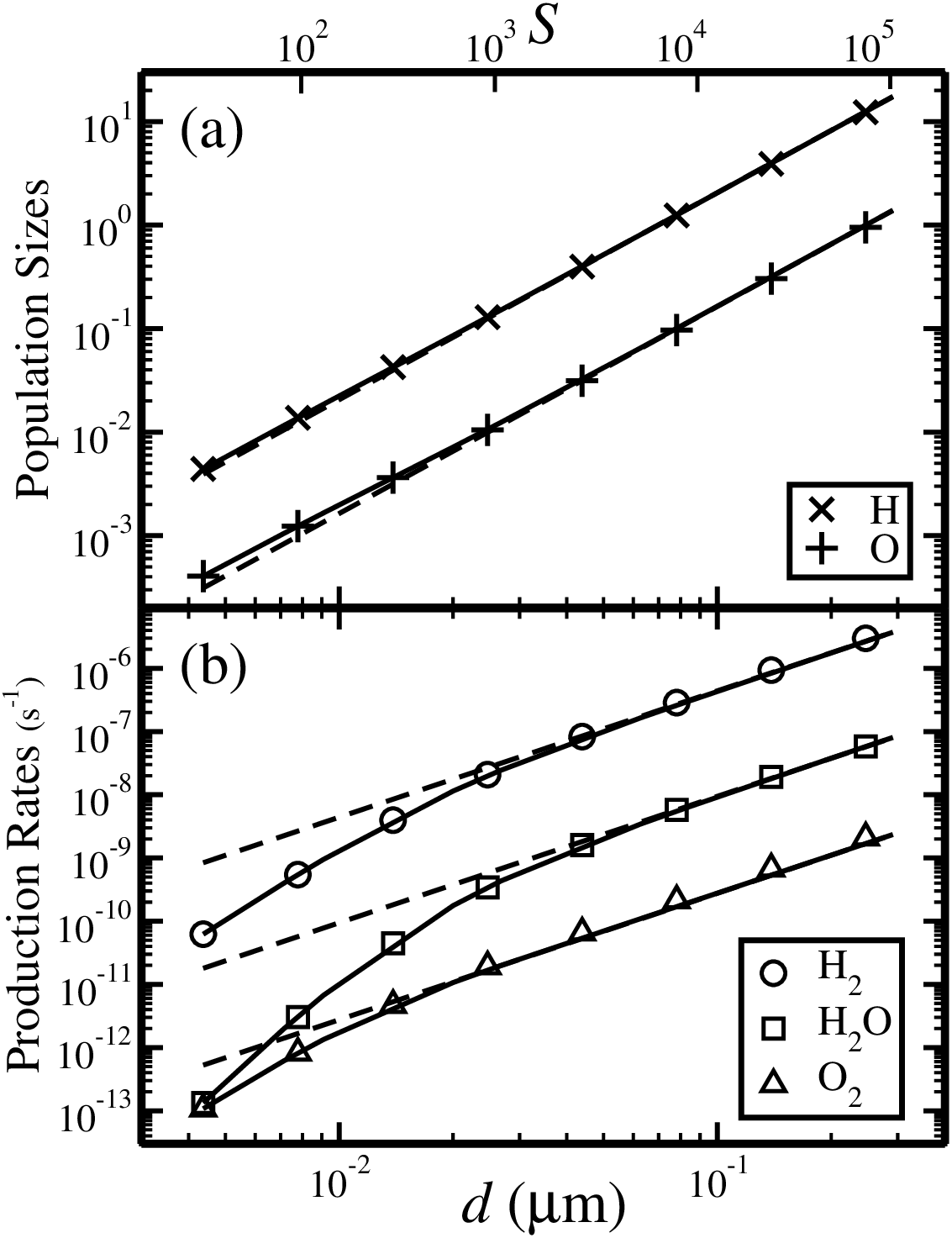}
\caption{
(a)
The population sizes of H ($\times$) and O ($+$) atoms
on a grain vs. grain diameter, $d$,
obtained from the moment equations
for the reaction network presented in 
Fig. 3(a).
The results are 
in excellent agreement with the 
master equation (solid line). 
The rate equation results are also shown (dashed line). 
(b) 
The production rates 
of H$_2$ (circles),
H$_2$O (squares)
and O$_2$ (triangles)
per grain
vs. $d$, 
obtained by the moment equations.
The results are 
in excellent agreement with the 
master equation (solid line). 
In the limit of large grains they also coincide with
the rate equations (dashed line). 
Note that the moment equations are valid both in the limit of small population sizes 
(when $d$ is small), and even when the population is high (large $d$s).
The number of adsorption sites, 
$S$,
is also displayed.  
}
\label{fig5}
\end{figure}

\begin{figure}
\includegraphics[width=4.5in]{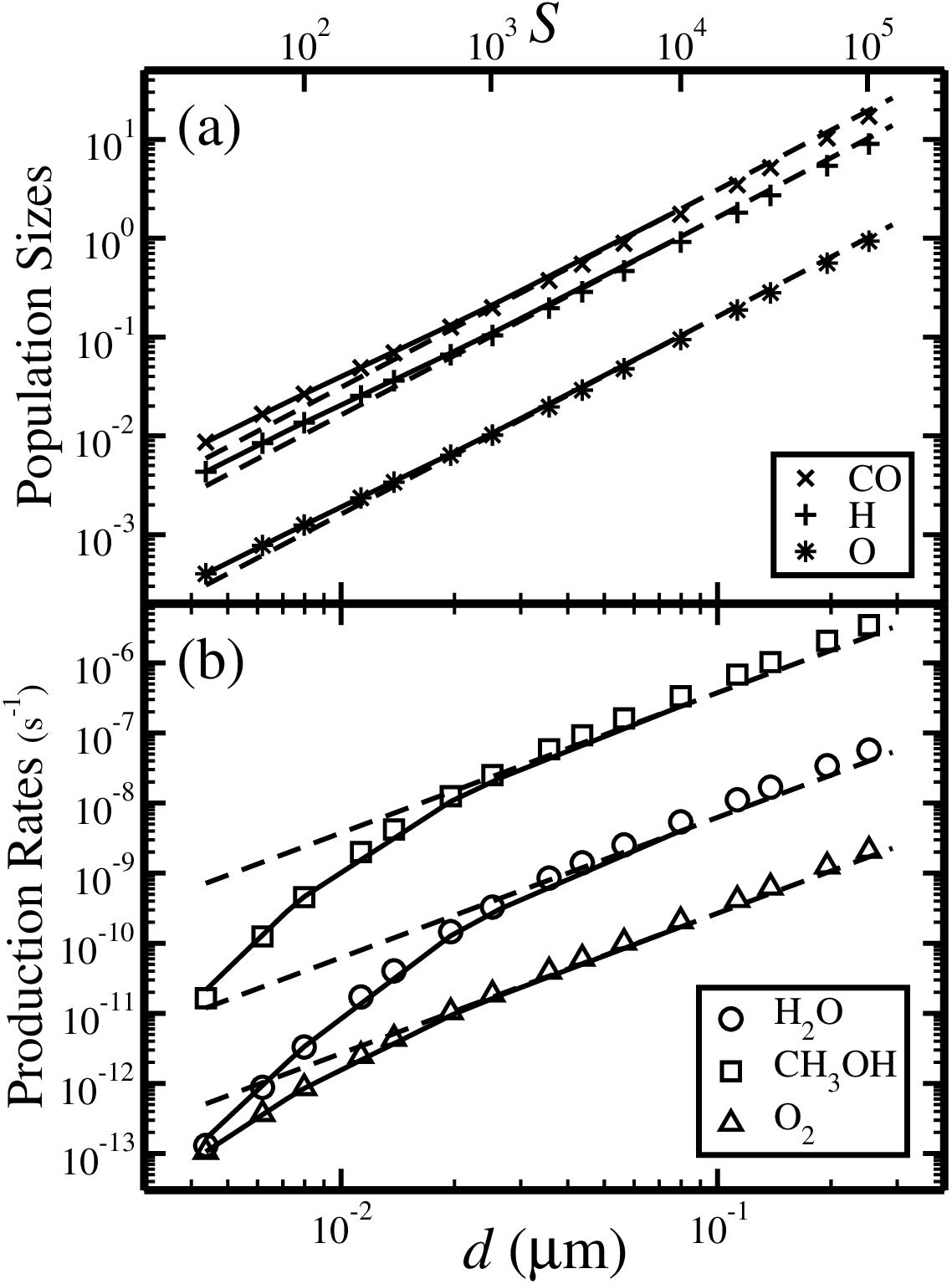}
\caption{
(a)
The population sizes of H ($+$), O ($\ast$) atoms
and CO molecules ($\times$)
on a grain vs. grain diameter, $d$,
obtained from the moment equations
for the reaction network presented in 
Fig. 3(b).
The results are 
in excellent agreement with the 
master equation (solid line). 
The rate equation results are also shown (dashed line). 
(b)
The production rates
of H$_2$O (circles),  
CH$_3$OH (squares) 
and O$_2$ (triangles) molecules 
on a grain vs. $d$, obtained from the moment equations.
The results are in agreement with the master
equation (solid line).
For small grains, the rate equation results
(dashed line)
show significant deviations. 
}
\label{fig6}
\end{figure}

Let us now examine how the moment equation method performs 
in a more complex network.
Recent laboratory experiments provide evidence
that methanol in interstellar clouds is formed via reaction networks
on dust grains
\cite{Geppert2006,Watanabe2005}.
To simulate these networks,
consider the case in which a flux of CO molecules
is added to the network shown in Fig. 
\ref{fig4}(a).
This gives rise to the 
network shown in 
Fig.
\ref{fig4}(b), 
which includes the
following sequence of hydrogen addition
reactions
\cite{Stantcheva2002}:
H + CO $\rightarrow$ HCO 
($X_1 + X_7 \rightarrow X_8$),
H + HCO $\rightarrow$ H$_2$CO 
($X_1 + X_8 \rightarrow X_9$),
H + H$_2$CO $\rightarrow$ 
H$_3$CO ($X_1 + X_9 \rightarrow X_{10}$)
and
H + H$_3$CO $\rightarrow$ 
CH$_3$OH ($X_1 + X_{10} \rightarrow X_{11}$).
Two other reactions that involve oxygen atoms also take place:
O + CO $\rightarrow$ CO$_2$ ($X_2 + X_7 \rightarrow X_{12}$)
and 
O + HCO $\rightarrow$ CO$_2$ + H 
($X_2 + X_8 \rightarrow X_{12} + X_1$).
This network was studied before using the multiplane method,
which required about a thousand equations compared to about a
million equations in the master equation with similar cutoffs. 
The moment equations include one equation for each species 
(node in the graph) and
one equation for each reaction (edge in the graph), 
namely, the network shown in 
Fig.
\ref{fig4}(b) requires only 17 equations.
We have performed extensive simulations of this network 
using the moment equations and found that they are in
agreement with the master equation results. 
In Fig. \ref{fig6}(a) we present 
the population sizes of
H ($+$),
O ($\ast$)
and 
CO ($\times$)
on a grain vs. grain diameter,
as obtained from the moment equations.
In Fig. \ref{fig6}(b) we present 
the results of the moment equations
for the formation rates of CH$_3$OH (squares),
H$_2$O (circles)
and O$_2$ (triangles)  
and compare them to 
the master equation (solid line) 
and the rate equations (dashed lines).
As before, the moment equations prove to be consistent 
with the master equation,
and for large grains with the rate equations as well.  
In the simulations presented here
the activation energies for diffusion and desorption
of CO, HCO, H$_2$CO and H$_3$CO
were taken as
$E_0(7) = 50 $,
$E_1(7) = 55 $,
$E_0(8) = 52 $,
$E_1(8) = 58 $,
$E_0(9) = 53 $,
$E_1(9) = 59 $
and
$E_0(10) = 55 $,
$E_1(10) = 62 $ meV,
respectively.
The flux of CO molecules was taken as
$F_7 = 0.2 F_1$.

In complex reaction networks the reduction in the number of equations
becomes extremely significant.
For each reactive species, $X_i$ (denoted by a node in the graph), 
there is one equation for the time derivative of the first moment
$\langle N_i \rangle$.
For each reaction, between species $X_i$ and $X_j$ (represented by an edge),
there is one equation for the time derivative of the second moment
$\langle N_iN_j \rangle$.
The number of equations is thus
$N_E^{\rm moment} = N_{\rm n} + N_{\rm e}$, where 
$N_{\rm n}$ and $N_{\rm e}$
are the numbers of nodes and edges, respectively.
Most reaction networks are sparse, namely the number of edges is 
of the same order of magnitude as the number of nodes.
As a result, the number of moment equations is roughly {\it linearly} 
dependent on the number of species. 
This is in contrast with the exponential growth of the master equation
and the polynomial growth of the multiplane equations
\cite{Lipshtat2004}.

\section{Diagrammatic Construction of the Equations}
\label{sec:Diagrams}

In the previous Section we described a procedure for the construction
of the moment equations, that 
consists of three steps.
In the first step the desired set of moment equations is determined
according to the network graph. This set includes one equation for
the first moment associated with each species (node) and one equation
for the second moment associated with each reaction (edge or loop).
In the second step the time derivative of each of these 
moments is expressed as a suitable linear combination of 
first, second and third moments.
In the third step the equations are brought into a closed form, by
expressing all third moments in terms of first and second moments
according to 
Eq. (\ref{eq:barzel_rules}).
This procedure is suitable for simple networks but becomes difficult
to execute as the network complexity increases.

Here we present a simple and highly efficient diagrammatic
approach for the construction of the moment equations
for any given network.
Consider the reaction networks shown in Fig.
\ref{fig4}. 
Each node in the graphs represents one of the reactive species, 
and each edge stands for a reaction between two species. 
A loop represents a reaction between two atoms/molecules 
of the same species.
The list of moments that should be included in the moment equations
is easily obtained from the graph:
(a) For each each node, $i$, there is an equation for the 
first moment $\langle N_i \rangle$;
(b) For each edge connecting the nodes $i$ and $j$
there is an equation for the second moment $\langle N_iN_j \rangle$; 
(c) for each loop attached to node $i$ there is an equation for
the second moment $\langle N_i^2 \rangle$. 

To derive the moment equations from the graph of a given network,
we show how each element in the graph is 
expressed in the master equation and then translate it into
a corresponding term in the moment equations.
Consider first the contribution of nodes. 
Each node represents a species, that may  
be adsorbed on the surface or desorbed from it.
In the master equation, these two processes are
described by

\newcommand{\vertex}[2][0.4]
{
\setlength{\unitlength}{#1cm}
\begin{picture}(2,0)(0,0.8)
\thicklines
\put(1,1){\circle{1.5}}
\put(0,0){\makebox(2,2){$#2$}}
\end{picture}
}

\newcommand{\edge}[3][0.4]
{
\setlength{\unitlength}{#1cm}
\begin{picture}(1.7,0)(0,-0.5)
\thicklines
\put(1,0.975){\circle{1.5}}
\put(0,0.025){\makebox(2,2){$#2$}}
\put(1,-1.875){\circle{1.5}}
\put(0,-2.875){\makebox(2,2){$#3$}}
\put(1,0.225){\line(0,-1){1.35}}
\end{picture}
}

\newcommand{\edgepro}[4][0.4]
{
\setlength{\unitlength}{#1cm}
\begin{picture}(2.1,0)(0,-0.5)
\thicklines
\put(1,0.975){\circle{1.5}}
\put(0,0.025){\makebox(2,2){$#2$}}
\put(1,-1.875){\circle{1.5}}
\put(0,-2.875){\makebox(2,2){$#3$}}
\put(1,0.225){\line(0,-1){1.35}}
\put(1,-1.124){\makebox(1,1.351){$#4$}}
\end{picture}
}

\newcommand{\Hedge}[4][0.4]
{
\setlength{\unitlength}{#1cm}
\begin{picture}(3,0)(0,-1.5)
\thicklines
\put(0,-1){\circle{1.5}}
\put(-1,-1.95){\makebox(2,2){$#2$}}
\put(3,-1){\circle{1.5}}
\put(2,-1.95){\makebox(2,2){$#3$}}
\put(0.75,-1){\line(1,0){1.5}}
\put(0.75,-0.0){\makebox(1.7,-0.4){$#4$}}
\end{picture}
}

\newcommand{\ozen}[3][0.4]
{
\setlength{\unitlength}{#1cm}
\begin{picture}(2.1,0)(0,#2)
\thicklines
\qbezier(0.6,-1.275)(-0.4,1.4)(1,1.475)
\qbezier(1.4,-1.275)(2.4,1.4)(1,1.475)
\put(1,-1.875){\circle{1.5}}
\put(0,-2.875){\makebox(2,2){$#3$}}
\end{picture}
}

\newcommand{\ozenpro}[3][0.4]
{
\setlength{\unitlength}{#1cm}
\begin{picture}(2.1,0)(0,0)
\thicklines
\qbezier(0.6,-1.275)(-0.4,1.4)(1,1.475)
\qbezier(1.4,-1.275)(2.4,1.4)(1,1.475)
\put(2.1,0.2){\makebox(2,2)[l]{$#3$}}
\put(1,-1.875){\circle{1.5}}
\put(0,-2.875){\makebox(2,2){$#2$}}
\end{picture}
}

\begin{eqnarray}
\vertex{i} &\longrightarrow & 
F_i[P(\dots ,N_i-1, \dots)-P(\dots ,N_i, \dots)]
\label{eq:node_master}
\\
&+&W_i[(N_i+1)P(\dots ,N_i+1, \dots)-
N_iP(\dots ,N_i, \dots)].
\nonumber 
\end{eqnarray}

\noindent 
To construct the equation for  
$\dot{\langle N_i \rangle}$
we sum over all the terms of the master equation,
according to 
Eq. (\ref{eq:moment_abc}). 
The terms above then make the following contribution  
to the moment equations

\begin{equation}
\vertex{i} \longrightarrow  F_i - W_i \langle N_i \rangle.
\label{eq:node_moment_1}
\end{equation}

\noindent 
In case that the species $i$ reacts with itself,
one needs to include 
an equation for $\dot{\langle N_i^2 \rangle}$.
As before, a proper summation over the master equation is performed,
but here
the contribution of the node $i$
will be

\begin{equation}
\vertex{i} \longrightarrow  F_i+
2F_i \langle N_i \rangle -2W_i \langle N_i^2 \rangle
+W_i \langle N_i \rangle.
\label{eq:node_moment_2ozen}
\end{equation}

\noindent 
If the species $i$ reacts with some other species, $j$,
it will be necessary to include an equation for 
$\dot{\langle N_iN_j \rangle}$.
The contribution of the node $i$ to this equation
is 

\begin{equation}
\vertex{i} \longrightarrow  F_i \langle N_j \rangle 
-W_i \langle N_iN_j \rangle.
\label{eq:node_moment_2edge}
\end{equation}

\noindent 
Note that in the equation for 
$\dot{\langle N_iN_j \rangle}$, 
one should also include the analogous term 
for the node $j$.

Consider the contribution of edges to the moment equations.
Each edge represents a reaction between two different species.
In the master equation, the term related to the edge between
$i$ and $j$ is 

\begin{eqnarray}
\edge{i}{j} & \longrightarrow & 
(A_i+A_j)[(N_i+1)(N_j+1)P(\dots N_i+1,N_j+1 \dots)
- N_iN_jP(\dots N_i,N_j \dots)].
\nonumber \\
\end{eqnarray}

\noindent
The contributions of this term to the moment equations 
appear in all the 
equations for moments involving either 
species $i$ or species $j$.
For example, the contribution to the equation for 
$\dot{\langle N_iN_j \rangle}$ 
is

\begin{eqnarray}
\edge{i}{j} & \longrightarrow & -(A_i+A_j)(\langle N_i^2N_j \rangle + 
\langle N_iN_j^2 \rangle 
- \langle N_iN_j \rangle).
\nonumber \\
\end{eqnarray}

\noindent
After applying the rules of 
Eq. (\ref{eq:barzel_rules}),
this term simplifies to 
$-(A_i+A_j)\langle N_iN_j \rangle$.

In general, each moment equation is obtained by a proper summation
over terms of the master equation, using
Eq. (\ref{eq:moment_abc}).
Each term in the master equation describes a certain process such
as adsorption, desorption or reaction.
Consider the equations for 
$\dot{\langle N_i \rangle}$
and
$\dot{\langle N_i^2 \rangle}$.
When the summations are carried out, 
the only non-vanishing contributions
are from processes that involve the species $X_i$. 
Similarly,
consider the equation for
$\dot{\langle N_iN_j \rangle}$. 
In this case the only non-vanishing contributions are
from processes that involve either the species $X_i$
or the species $X_j$.
This gives rise to a dramatic simplification in the construction
of the moment equations.
In the equation for any desired moment 
we need to consider only those master equation terms 
related to the species that
appear in that particular moment.
In the diagrams, this means
including only graph elements
that are directly connected to 
the corresponding nodes.
The complete set of translation rules from a given graph element
to the corresponding terms in the moment equations is given
in Tables 
\ref{tab:moment_terms1} - 
\ref{tab:moment_terms3}. 

To demonstrate the construction of the moment equations 
we reconsider the reaction network shown in 
Fig. \ref{fig4}(a).
First, we identify the relevant moments
associated with the nodes, edges and loops in the network graph.
In this particular network, the nodes call for
$\langle N_1 \rangle$, 
$\langle N_2 \rangle$ 
and 
$\langle N_3 \rangle$.
The edges call for 
$\langle N_1N_2 \rangle$ 
and 
$\langle N_1N_3 \rangle$.
Finally, the two loops in the graph 
require the inclusion 
of 
$\langle N_1^2 \rangle$ 
and 
$\langle N_2^2 \rangle$.
Altogether, there are seven equations.
Let us begin with the equation for 
$\dot{\langle N_1 \rangle}$.
This species is represented by a node 
connected to two edges and one loop.
Using our diagrammatic scheme, this equation
takes the form

\begin{eqnarray}
\label{eq:diag1}
\dot{\langle N_1 \rangle} &=& \vertex{1} +
\edge{1}{2} + \edge{1}{3} + \ozen{-0.5}{1}.
\\
\nonumber 
\end{eqnarray}

\noindent
This equation simply consists of all the graph elements related
to the species $X_1$.
That diagrammatic sum can be translated, 
using 
Table \ref{tab:moment_terms1},
giving rise to the equation

\begin{eqnarray}
\dot{\langle N_{1} \rangle} &=&
F_{1} - W_{1} \langle N_{1} \rangle 
-(A_{1}+A_{2}) \langle N_{1}N_{2} \rangle \nonumber \\ 
&-&(A_{1}+A_{3}) \langle N_{1}N_{3} \rangle 
-2A_{1} (\langle N_{1}^2 \rangle - \langle N_{1} \rangle)  
\label{eq:moment_OH_1}
\end{eqnarray}

\noindent
The same diagrammatic equation
[Eq. (\ref{eq:diag1})]
is used when writing the equation
for the time derivative of
$\langle N_1^2 \rangle$.
However, the translation rules from graph elements to
equation terms are different and are given in
Table \ref{tab:moment_terms2}.
Using these rules we obtain

\begin{eqnarray}
\dot{\langle N_{1}^2 \rangle} &=&
F_{1}+(2F_{1}+W_{1}+4A_{1}) \langle N_{1} \rangle 
-(2W_{1} + 4A_{1}) \langle N_{1}^2 \rangle   
\nonumber \\
&-&(A_{1}+A_{2}) \langle N_{1}N_{2} \rangle 
-(A_{1}+A_{3}) \langle N_{1}N_{3} \rangle.
\label{eq:moment_OH_2}
\end{eqnarray}

\noindent
Next, we construct the moment equation for 
$\dot{\langle N_1N_2 \rangle}$.
This time we need to include some 
additional graph elements, 
namely, those related to $X_2$, 
so the diagrammatic equation 
takes the form  

\begin{eqnarray}
\dot{\langle N_1N_2 \rangle} &=&
\vertex[0.3]{1} + \vertex[0.3]{2} + 
\edge[0.3]{1}{2} + \edge[0.3]{1}{3} + 
\ozen[0.3]{-0.5}{1} + \ozen[0.3]{-0.5}{2}. 
\\
\nonumber 
\end{eqnarray}

\noindent
This is an equation for a mixed second moment,
that should be translated according to Table
\ref{tab:moment_terms3}. 
The resulting equation is

\begin{equation}
\dot{\langle N_{1} N_{2} \rangle} =
F_{1} \langle N_{2} \rangle + F_2 \langle N_1 \rangle
-(W_{1} + W_2) \langle N_{1} N_2 \rangle 
-(A_{1}+A_{2}) \langle N_{1}N_{2} \rangle.  
\label{eq:moment_OH_3}
\end{equation}

\noindent
The last equation that we  
demonstrate explicitly is for the moment
$\dot{\langle N_3 \rangle}$. 
This time the diagrammatic equation
will include an additional graph element
stating the fact that $X_3$ 
is produced by a reaction between the other two species
in the network.
The diagrammatic equation is

\begin{eqnarray}
\dot{\langle N_3 \rangle} =
\vertex{3} + \edge{1}{3} + \edgepro{1}{2}{3},
\\
\nonumber
\end{eqnarray}

\noindent
which, 
according to 
Table \ref{tab:moment_terms1} 
translates to

\begin{eqnarray}
\dot{\langle N_{3} \rangle} &=&
F_{3} - W_{3} \langle N_{3} \rangle 
- (A_{1}+A_{3}) \langle N_{1}N_{3} \rangle  
+(A_{1}+A_{2}) \langle N_{1}N_{2} \rangle. 
\label{eq:moment_OH_4}
\end{eqnarray}

This procedure enables an efficient and straightforward
construction of the set of moment equations for any given
graph, regardless of its complexity.
The construction process can be easily automated into a computer
program that receives the network structure as an input and provides
the set of equations as an output.
This opens the way to efficient stochastic modeling of complex reaction
networks of any size.

\section{Summary}
\label{Summary}

We have presented an efficient method for the stochastic simulation
of complex reaction networks. The method is based on moment 
equations.
It accounts correctly for the 
reaction rates not only for macroscopic populations,
where rate equations apply, but also in the limit of low copy numbers
where fluctuations dominate. 
The method exhibits several crucial advantages over existing 
techniques for stochastic simulations.
The number of equations is reduced to the absolute minimum in 
a stochastic simulation, namely one equation for each 
reactive species and one equation for each reaction. 
For any given network the set of moment equations can 
be constructed easily and efficiently
using a diagrammatic approach.
The equations are linear, thus, for steady-state
conditions they can be easily solved using algebraic methods.

We have demonstrated the method for
reactions taking place on dust-grain surfaces 
in interstellar clouds.
We have shown that it
applies very well 
even under the extreme interstellar conditions of low gas density 
and sub-micron grain sizes.
The moment equations can be easily coupled to the rate equation models
of interstellar gas-phase chemistry.
Therefore, the proposed method is expected to enable the incorporation
of grain-surface chemistry into these models.

We expect this method to be useful for the simulation of 
many other systems that exhibit a related 
mathematical structure
\cite{Biham2005}.
One such application is for reactions taking place on terrestrial
surfaces such as metal catalysts in nanometric systems
\cite{Suchorski1998,Suchorski1999,Suchorski2001,Johanek2004,Pineda2006,Liu2002}. 
Another example is
aerosol chemistry in stratospheric clouds,
where complex reaction networks involving aerosol
particles take place
\cite{Hanson1994}.
The gas density is not nearly as low as in the
interstellar clouds.
However, the small size of the aerosol particles
and the low density of some of the reactive species
may require stochastic methods.
Another example is genetic networks in cells and bacteria 
\cite{McAdams1997,Paulsson2000,Lipshtat2006,Alon2006}.
These networks describe the process of protein synthesis and
its regulation. They include the interactions between genes, mRNA's
and proteins. Since some of these components appear in low copy 
numbers, the simulation of these networks requires stochastic methods. 

We thank Azi Lipshtat for helpful discussions.
This work was supported by the Israel Science Foundation and the
Adler Foundation for Space Research.

\newpage
\clearpage

%
%

\begin{table}
\caption{
The moment equations method is valid for producing the first and second moments
of the distribution
$P(N)$
both for large and small grains.
These moments enable the evaluation of the population size of atoms per grain,
and the production rate of
H$_2$
molecules on a grain.
The one exception is for producing the first moment for large grains in the regime of first order kinetics.
In that case, one may simply use the rate equation model, which is valid for large grains,
or the corrected moment equations which always apply for large grains.}
\label{table1}
\end{table}

\begin{tabular}{||c|c||c|c||}
\cline{3-4} \cline{3-4}
\multicolumn{2}{c||}{} & \multicolumn{2}{c||}{
\bfseries The Validity of the Moment Equations}
\\ \cline{3-4} \cline{3-4}
\multicolumn{2}{c||}{} &
\multicolumn{1}{c|}{  Small Grain} &
\multicolumn{1}{c||}{ Large Grain}
\\ \hline \hline
First order    & $\N$    & $\surd$     & $\times$
\\ 
               & $\N[2]$ & $\surd$     & $\surd$
\\ \hline
Second order   & $\N$    & $\surd$     & $\surd$
\\ 
               & $\N[2]$ & $\surd$     & $\surd$
\\ \hline \hline
\end{tabular}

\newpage
\clearpage

%
%

\begin{table}
\caption{
Diagrammatic approach for the construction of the moment equations.
The equation for 
$\dot{\langle N_i \rangle}$ is given by the sum of all graph elements 
related to the species $X_i$.
To construct this moment equation,
each graph element is substituted by the Reduced Equation Term,
obtained using the rules of Eq. (\ref{eq:barzel_rules}). 
The middle column provides the exact terms of the moment equation,
before the approximation of 
Eq. (\ref{eq:barzel_rules}) is applied. 
}
\label{tab:moment_terms1}
\end{table}

\newcommand{\rb}[2][-2.5]{\raisebox{#1ex}{#2}}

\begin{tabular}{||c|c|c||}
\hline \hline
\multicolumn{3}{||c||}{\rule[-1.51mm]{0mm}{6mm}
{\bfseries Graph Elements Translation for the 
$\dot{\langle N_i \rangle}$ Equation }}
\\ \hline
Graph Element           &  Equation Term                
                       & Reduced Equation Term
\\ \hline \hline        & & \\
$\vertex{i}$            & $ F_i-W_i \langle N_i \rangle$          
   & $F_i-W_i \langle N_i \rangle$
\\ & & \\ \hline        & & \\ 
$\ozen[0.35]{-0.5}{i}$  & $-2A_i${\Large (}$\langle N_i^2 \rangle 
- \langle N_i \rangle${\Large )}& $-2A_i${\Large (}$\langle N_i^2 \rangle 
- \langle N_i \rangle${\Large )} 
\\ & & \\ & & \\ \hline & & \\
$\Hedge{i}{j}{}$        & $-(A_i+A_j)\langle N_iN_j \rangle$      
                & $-(A_i+A_j)\langle N_iN_j \rangle$ 
\\ & & \\  \hline       & & \\
$\Hedge{k}{j}{i}$       & $(A_k+A_j)\langle N_kN_j \rangle$        
              & $(A_k+A_j)\langle N_kN_j \rangle$ 
\\ & & \\  \hline       & & \\
$\ozenpro[0.35]{k}{i}$  & $A_k${\Large (}$\langle N_k^2 \rangle 
- \langle N_k \rangle${\Large )}    & $A_k${\Large (}$\langle N_k^2 \rangle 
- \langle N_k \rangle${\Large )}
\\ & & \\ & & 
\\ \hline \hline 
\end{tabular}

\newpage
\clearpage

%
%

\begin{table}
\caption{
The equation for 
$\dot{\langle N_i^2 \rangle}$ is the sum of all graph elements 
related to the species $X_i$.
Some of the exact equation terms, given in the middle column,
include third order moments. 
The reduced equation terms, given in the third column,
consist of only first and second order moments.
}
\label{tab:moment_terms2}
\end{table}

\begin{tabular}{||c|c|c||}
\hline \hline
\multicolumn{3}{||c||}{\rule[-1.51mm]{0mm}{6mm}{\bfseries 
Graph Elements Translation for the $\dot{\langle N_i^2 \rangle}$ Equation }}
\\ \hline
Graph Element           &  Equation Term                                
              & Reduced Equation Term
\\ \hline \hline        
                        & \rb{$F_i+2F_i\langle N_i \rangle$}               
           & \rb{$F_i+2F_i\langle N_i \rangle$} 
\\
$\vertex{i}$            & \rb{$+W_i \langle N_i \rangle -2W_i \langle N_i^2 \rangle$}
 & \rb{$+W_i \langle N_i \rangle -2W_i \langle N_i^2 \rangle$}
\\ & & \\ \hline        
                        & \rb{$-4A_i${\Large 
(}$\langle N_i^3 \rangle -2\langle N_i^2 \rangle$}   & \\ 
$\ozen[0.35]{-1}{i}$    & \rb{$+\langle N_i \rangle${\Large )}}                  
                 & $-4A_i${\Large (}$\langle N_i^2 \rangle - \langle N_i \rangle${\Large )} 
\\ & & \\ \hline         
                        & \rb{$(A_i+A_j)${\Large (}$\langle N_iN_j \rangle$}         
             & \\
$\Hedge{i}{j}{}$        & \rb{$-2\langle N_i^2N_j \rangle${\Large )}}                   
          & $-(A_i+A_j)\langle N_iN_j \rangle$ 
\\ & & \\  \hline      
                        & \rb{$(A_j+A_k)${\Large (}$2\langle N_iN_jN_k \rangle$}        
          & \\
\rb[-1]{$\Hedge{j}{k}{i}$}  & \rb{$+\langle N_jN_k \rangle${\Large )}}                
                & $(A_j+A_k) \langle N_jN_k \rangle$
\\ & & \\  \hline  
\rb{$\ozenpro[0.35]{k}{i}$} & \rb{$A_k${\Large
 (}$2\langle N_iN_k^2 \rangle - 2\langle N_iN_k \rangle$} & \\ 
                            & \rb{$+\langle N_k^2 \rangle -
 \langle N_k \rangle${\Large )}}             & $A_k${\Large (}$\langle N_k^2 \rangle 
- \langle N_k \rangle${\Large )}
\\ & & \\ \hline \hline
\end{tabular}  

\newpage
\clearpage

%
%

\begin{table}
\caption{
The equation for 
$\dot{\langle N_iN_j \rangle}$ is the sum of all graph elements 
related to the species $X_i$ and $X_j$.
Elements related to both species should be included only once.
}
\label{tab:moment_terms3}
\end{table}

\begin{tabular}{||c|c|c||}
\hline \hline
\multicolumn{3}{||c||}{\rule[-1.51mm]{0mm}{6mm}{\bfseries 
Graph Elements Translation for the $\dot{\langle N_iN_j \rangle}$ Equation }}
\\ \hline
Graph Element           &  Equation Term                                             
 & Reduced Equation Term
\\ \hline \hline        
                        & \rb{$F_i\langle N_j \rangle 
+F_j\langle N_i \rangle$}         
 & \rb{$F_i\langle N_j \rangle +F_j\langle N_i \rangle$}
\\
$\vertex{i}+\vertex{j}$   & \rb{$-(W_i+W_j) \langle N_iN_j \rangle$}                   
  & \rb{$-(W_i+W_j) \langle N_iN_j \rangle$}
\\ & & \\ \hline                                             
& \rb{$-2A_i ( \langle N_i^2N_j  \rangle - \langle N_iN_j \rangle)$}   
&  \\ 
$\ozen[0.35]{-1}{i}$ {\Huge (}$\ozen[0.35]{-1}{j}$ {\Huge )} 
& \rb{{\Large (}$-2A_j (\langle N_iN_j^2\rangle - \langle N_iN_j \rangle)${\Large )}} 
                                                                                        
 & {$0$} 
\\ & & 
\\ \hline        & \rb{$-(A_i+A_j)${\Large (}$\langle 
N_i^2N_j \rangle$}                 
    & \\
$\Hedge{i}{j}{}$        & \rb{$+\langle N_iN_j^2 \rangle 
- \langle N_iN_j \rangle${\Large )}}  
     & $-(A_i+A_j)\langle N_iN_j \rangle$ 
\\ & & \\  \hline      
\rb{$\Hedge{i}{k}{}$}                     
& \rb{$-(A_i+A_k)\langle N_iN_jN_k \rangle$}         
                       & \\
\rb{{\huge ( }$\Hedge{j}{k}{}$\hspace{1mm} {\huge )}} 
& \rb{{\Large (}$-(A_j+A_k)\langle N_iN_jN_k \rangle${\Large )}}      & $0$
\\ & & \\  \hline       
                                                             
& $-(A_i+A_k)${\Large (}$\langle N_iN_jN_k \rangle$                  &     \\
{$\Hedge{i}{k}{j}$}                                          
& $-\langle N_i^2N_k \rangle + \langle N_iN_k \rangle${\Large )}       &     \\
\rb[-1.9]{{\huge ( }$\Hedge{j}{k}{i}$\hspace{1mm} {\huge )}} 
&\rb[-1.9]{{\Large (}$
-(A_j+A_k)${\Large (}$\langle N_iN_jN_k \rangle$}      & $0$ \\
                                                       
& \rb[4]{$-\langle N_j^2N_k \rangle 
+ \langle N_jN_k \rangle${\Large )}{\Large )}} &     \\
\hline
\rb{$\Hedge{k}{m}{i(j)}$}    &\rb{$-(A_k+A_m)\langle N_kN_mN_{j(i)}
 \rangle$} & \rb{$0$} 
\\ & & \\ \hline & & \\
$\ozenpro[0.35]{k}{i(j)}$  & $A_k(\langle N_k^2N_{j(i)} \rangle 
- \langle N_kN_{j(i)} \rangle)$    & $0$
\\ & & \\ & & \\ \hline \hline
\end{tabular}

\newpage
\clearpage 
 
%
%

%
%

%
%

%
%

%
%

%
%

\end{document}